\documentclass[twocolappendix]{emulateapj}

\usepackage[usenames,dvipsnames]{color}

\usepackage{comment}
\usepackage{soul,color,amsmath}
\usepackage{graphicx}

\newenvironment{rcases}
  {\left.\begin{aligned}}
  {\end{aligned}\right\rbrace}
  
\shorttitle{Eccentric Jupiters} 
\shortauthors{Duffell \& Chiang}

\begin{document}

\title{Eccentric Jupiters via Disk-Planet Interactions}

\author{Paul C. Duffell and Eugene Chiang}
\affil{Department of Astronomy and Theoretical Astrophysics Center, University of California, Berkeley}
\email{duffell@berkeley.edu, echiang@astro.berkeley.edu}

\begin{abstract}
Numerical hydrodynamics calculations are performed
to determine conditions under which giant
planet eccentricities can be excited by
parent gas disks. Unlike in other studies,
Jupiter-mass planets are found to have their
eccentricities amplified --- provided their
orbits start eccentric.
We disentangle the web of co-rotation, co-orbital,
and external resonances to show that
this finite-amplitude instability 
is consistent with that predicted analytically.
Ellipticities can grow until they reach
of order the disk's aspect ratio, beyond which
the external Lindblad resonances that excite
eccentricity are weakened
by the planet's increasingly supersonic
epicyclic motion. Forcing the planet to still
larger eccentricities causes catastrophic
eccentricity damping as the planet
collides into gap walls.
For standard parameters,
the range of eccentricities for instability
is modest; the threshold eccentricity for
growth ($\sim$$0.04$) is not much smaller than the final
eccentricity to which orbits grow ($\sim$$0.07$).
If this threshold eccentricity can be lowered (perhaps by
non-barotropic effects), and if the eccentricity driving
documented here survives in 3D, it may robustly
explain the low-to-moderate eccentricities $\lesssim 0.1$
exhibited by many giant planets (including Jupiter and
Saturn), especially those without
planetary or stellar companions.
\end{abstract}


\section{Introduction}
\label{sec:intro}	

One of the most surprising revelations of
Doppler exoplanet surveys is the
prevalence of Jupiter-mass planets on
highly elliptical orbits
\citep[e.g.,][]{2005PThPS.158...24M}. 
At orbital periods $\gtrsim 10$ days,
beyond the reach of tidal circularization,
giant planet eccentricities span the full
gamut from near-zero to near-unity.
There is growing evidence that
gravitational interactions between
planets can explain the
extravagant eccentricities observed
\citep[e.g.,][]{2005ApJ...627.1001T, 2008ApJ...686..603J, 2011ApJ...735..109W, 2014Sci...346..212D}.

But are planet-planet interactions
the whole story?
After removing observational biases,
a substantial fraction of 
giant planets
have low-to-moderate eccentricities:
$\sim$28\% have $e < 0.05$ --- our solar
system gas giants belong to this cohort ---
and fully half have $e < 0.15$
(\citealt{2011MNRAS.410.1895Z},
see their Figure 11, bottom panel).
These statistics are drawn
from the single-planet catalog
of \cite{2006ApJ...646..505B}. Continued
Doppler monitoring has not changed
the single status of many of these
planets, particularly at semimajor
axes $> 1$ AU (Bryan et al., submitted).
For solitary giants having no stellar or
planetary perturbers
in sight, we look instead to their parent
gas disks to understand how their
ellipticities may have arisen.

Planet-disk interactions are mediated
by resonances, of which there are
as many kinds as there are terms
in the Fourier expansion of the planet's
potential. Some resonances damp 
eccentricity while others excite it. 
Goldreich \& Sari (\citeyear{2003ApJ...585.1024G}, hereafter GS03) outlined the circumstances
whereby certain resonant interactions
could dominate others to excite
eccentricity in the net. The planet
would need to (1) carve out a gap
around its orbit, and (2) have its
eccentricity exceed a threshold value,
which GS03 estimated to be on the order
of a few percent (see Sections 2 and
4.1 for the technical details).
Amplification of
eccentricity by disk torques could then
proceed, presumably until the planet crashed
into the gap walls. Eccentricities excited
by disks would then be limited by the
fractional radial widths of gaps, of order
$\sim$0.1. Planet-disk interactions can
thus be argued to be relevant for
eccentricities in the range $\sim$0.01--0.1.

The finite-amplitude instability
of GS03 has seen little if any support
in numerical studies. Overwhelmingly,
planets the mass of Jupiter or lower
are seen in numerical simulations
to have their eccentricities damped
\citep[e.g.,][]{2001A&A...366..263P, 
2006A&A...450..833C, 2007A&A...473..329C, 2010A&A...523A..30B, 2013MNRAS.428.3072D, 2013A&A...555A.124B}.
Some of these studies found
that eccentricities grow only
for planets of
relatively high mass
$\gtrsim 5$--$20 M_{\rm J}$, via a
mechanism that differs from the one
proposed by GS03 (see Section 2).
To our knowledge,
the one numerical study that reported
otherwise was by
\cite{2006ApJ...652.1698D}
who found that Jupiter-mass
planets could have their eccentricities
excited to values of $\sim$0.1.
It is unclear whether their results
vindicate the GS03 mechanism,
as \cite{2006ApJ...652.1698D} observed
eccentricities to grow starting from zero;
in other words,
no evidence was found for a
finite-amplitude instability.

Many previous numerical studies of
eccentricity evolution used a
``live-planet" approach: the planet's
orbit was free to evolve under the
action of disk torques.
Although natural enough, a live-planet
simulation can be tricky to diagnose
because all parameters are in flux.
We advocate here a ``fixed-planet"
methodology: the planet is kept on
a fixed eccentric orbit; the disk
is allowed to relax to a quasi-steady
state (one that oscillates consistently
with the epicyclic phase of the planet);
and disk forces on the planet are
then measured to extract the rate
of change of planet eccentricity $\dot{e}$
as a function of $e$.
The fixed-planet approach, as used
by, e.g., \citet{2010A&A...523A..30B}
and \citet{2013A&A...555A.124B},
permits greater control of environmental variables
and more systematic exploration of
parameter space.

In Section \ref{sec:theory} we
briefly review the
theoretical considerations underlying
how disks affect planetary
eccentricities. We summarize
the tenets of the theory of GS03 and
also itemize aspects of the problem
that they did not treat.
Section \ref{sec:numerics} describes
the numerical methods we employed to
measure $\dot{e}(e)$ for Jupiter-mass
planets. Results, including a head-to-head
comparison
with the predictions of GS03,
are given in Section \ref{sec:results}.
A summary and outlook is contained
in Section \ref{sec:discussion}.
The Appendix compiles all the 
formulae we used to test GS03,
drawn from several analytic studies.

\section{Theoretical Background}
\label{sec:theory}	

According to GS03, eccentricity excitation requires two ingredients:

\begin{itemize}

\item The planet must carve a deep enough
gap in the disk --- or more accurately, a
gap with steep enough density gradients ---
that first-order (as
expanded in the planet's eccentricity)
co-orbital Lindblad resonances situated at
the very gap center become weaker than
first-order external Lindblad resonances
located roughly a gas scale height $h$
away from gap center.
The latter drive eccentricity, while the
former damp eccentricity. Citing
calculations by \citet{1993ApJ...419..166A},
GS03 stated that the gap profile must be
such that the surface density
at the locations of the strongest
externals must be greater than the surface
density at gap center by at least a factor
of $\sim$3 for the externals to defeat the co-orbitals.\footnote{We will find in
practice that this requirement is met
only for gaps that are extremely deep
in the sense that their
central surface densities
are suppressed by about 3 orders
of magnitude relative to the
background disk.}

\item First-order co-rotation resonances,
which also damp eccentricity,
must be ``saturated" (weakened),
meaning that material librating
in co-rotation resonance must not be
replenished by viscous inflow of fresh
material \citep{2003ApJ...587..398O}.  
Saturation is effected for sufficiently
large $e > e_{\rm min}$; in other words,
eccentricity excitation is a
finite-amplitude instability.

\end{itemize}

As estimated by GS03, the minimum eccentricity
necessary for $\dot{e}>0$ is
\begin{equation}
e_{\rm min} \sim \left( \frac{w}{a} \right)^{5/3} \left( \frac{\nu}{\Omega_0 a^2} \right)^{2/3} q^{-1}
\label{eqn:emin}
\end{equation}
where $q$ is the planet-to-star mass ratio,
$a$ is the planet's semimajor axis,
$\Omega_0$ is the planet's
orbital frequency, $w$ is the gap width, and
$\nu$ is the kinematic
viscosity. To estimate $w$, GS03 balance the
one-sided principal (zeroth-order)
Lindblad torque with the local
viscous torque
\begin{equation} \label{eqn:cheat}
q^2 \Omega_0^2 \Sigma a^4 (a/w)^3 \sim \nu \Sigma a^2 \Omega_0
\end{equation}
and obtain
\begin{equation} \label{eq:width}
w/a \sim (q^2/\tilde\nu)^{1/3}
\end{equation}
where $\Sigma$ is
the disk surface density
and $\tilde \nu = \nu/(a^2 \Omega_0)$ is the
kinematic viscosity with dimensions scaled
out. Equation (\ref{eqn:cheat})
ignores changes in $\Sigma$ across
the gap which can actually be
substantial;\footnote{ \cite{2014ApJ...782...88F} 
did account for changes in surface
density when writing down (\ref{eqn:cheat}),
deriving a scaling relation for gap
depth that succeeds in reproducing
numerical results. These authors replaced
the left-hand $\Sigma$ with
$\Sigma_{\rm gap}$, the right-hand
$\Sigma$ with the unperturbed
value $\Sigma_0$, and $w$ with $h$
to arrive at a fairly accurate formula for
$\Sigma_{\rm gap}/\Sigma_0$.
Their argument and GS03's argument for
$w$ as presented here are not obviously
compatible.} we will, in any case,
test scaling relation (\ref{eq:width})
numerically in Section 4.3.
Substituting (\ref{eq:width})
into (\ref{eqn:emin}) gives
\begin{equation}
e_{\rm min} \sim (q \tilde \nu)^{1/9} \,.
\end{equation}
For our standard parameters of $q = 0.001$ and $\tilde \nu = 2.5 \times 10^{-6}$ (corresponding
to a Shakura-Sunyaev $\alpha = 0.002$ and disk aspect ratio $h/a = 0.036$),
equation (\ref{eqn:emin}) --- which
is not meant to be more
than an order-of-magnitude estimate ---
gives $e_{\rm min} \sim 0.1$.

For $e > e_{\rm min}$, a reasonable
expectation
not specifically discussed by GS03 is that
the eccentricity
should grow until the planet's radial
epicyclic motion causes it to collide
with the gap walls. The maximum eccentricity
$e_{\rm max}$ should then scale
as $w/a$.\footnote{We will present
evidence supporting this expectation.
Actually, we will find that
before $w/a$ comes into play,
the disk aspect ratio $h/a < w/a$ 
becomes relevant. See Section \ref{sec:anly}
on the supersonic weakening of Lindblad
resonances and how the weakening leads to
$e_{\rm max}$.}

A useful order-of-magnitude formula that
gives a sense of scale is the maximum rate
of eccentricity damping in the limit
of small $e$ and no
gap clearing \citep{1993ApJ...419..166A}:
\begin{equation}
\max |\dot e / e| \sim q \left( \frac{a}{h} \right)^4 { \Sigma_0 a^2 \over M_* } \Omega_0
\label{eqn:edote}
\end{equation}
where $\Sigma_0$ is the unperturbed
disk density. This maximum rate of
eccentricity change
is set by the co-orbital resonances.
Gap clearing
can only reduce the magnitude of
eccentricity changes (and potentially
change the sign).

Apsidal resonances are first-order
Lindblad resonances with pattern
speeds equal to the planet's apsidal
precession frequency (they have
wavenumbers $m=1$ and $\ell = 0$ in
the Fourier notation of \cite{1980ApJ...241..425G}).
They damp the planet's eccentricity \citep{1998AJ....116..489W, 2000prpl.conf.1135W}, but are argued by GS03 to be of modest
importance compared to $m > 1$ first-order Lindblad
resonances.\footnote{Apsidal (a.k.a.~secular)
torques will nevertheless be captured in our numerical
calculations. The disk eccentricity and apsidal profile
will relax to an equilibrium set by driving from
the eccentric planet and damping by viscosity;
the eccentric disk streamlines will 
backreact secularly onto the planet (probably damping
the planet's eccentricity). Although we will not
separate out the apsidal/secular torque, it is 
part of the total torque that we evaluate from the entire
disk; see equations
(\ref{eqn:torque})--(\ref{eqn:everything}).}

Other effects not covered by the linear theory of GS03 include torques exerted by material in the immediate vicinity
of the planet, on scales of order the Hill radius. Circumplanetary material (not necessarily bound
to the planet) may exert dynamical friction and strongly damp the planet's eccentricity.
Properly modeling circumplanetary flows is challenging and subject to numerical issues
such as how the planet's potential is smoothed and how accretion onto the planet is prescribed. Another nonlinear issue
concerns instabilities in deep gaps \citep{2009ApJ...690L..52L, 2010ApJ...712..198Y, 2013ApJ...769...41D, 2014ApJ...782...88F, 2015MNRAS.448..994K}.  If the planet mass is large enough or the viscosity is small enough, then gap walls can steepen to the point of triggering the
Rayleigh instability or the Rossby wave instability.  Gap walls can shed vortices that can stochastically
torque the planet.

Finally, we emphasize that the GS03
mechanism for 
eccentricity growth does not align with
the common view
that to drive eccentricity
requires near-brown dwarf masses and the dominant influence of the outer 1:3 Lindblad resonance.
\cite{2001A&A...366..263P}, \cite{2013MNRAS.428.3072D},
and \cite{2013A&A...555A.124B}
found eccentricity
driving only for relatively massive giants ($\gtrsim 5$--$20 M_{\rm J}$); these companions
opened such wide and deep gaps that they interacted primarily with their disks
via the outer 1:3 resonance, amplifying disk eccentricities which were then backreactively shared with the planet
by secular interactions (see also \citealt{2006A&A...447..369K}).
\cite{2013A&A...555A.124B} found eccentricity
growth for 5--10 $M_{\rm J}$ planets only when
such planets were forced to occupy substantially 
eccentric orbits, $e = 0.2$--0.4 (see their Figure 4).
We will find in the present study that the 1:3 resonance is not essential for eccentricity driving;
that it is possible to excite planetary eccentricities even for Jupiter-mass
planets,
starting with eccentricities as low as a few percent,
along the lines envisioned by GS03.

\section{Numerical Method}
\label{sec:numerics}	

\begin{figure*}
\epsscale{1.15}
\plotone{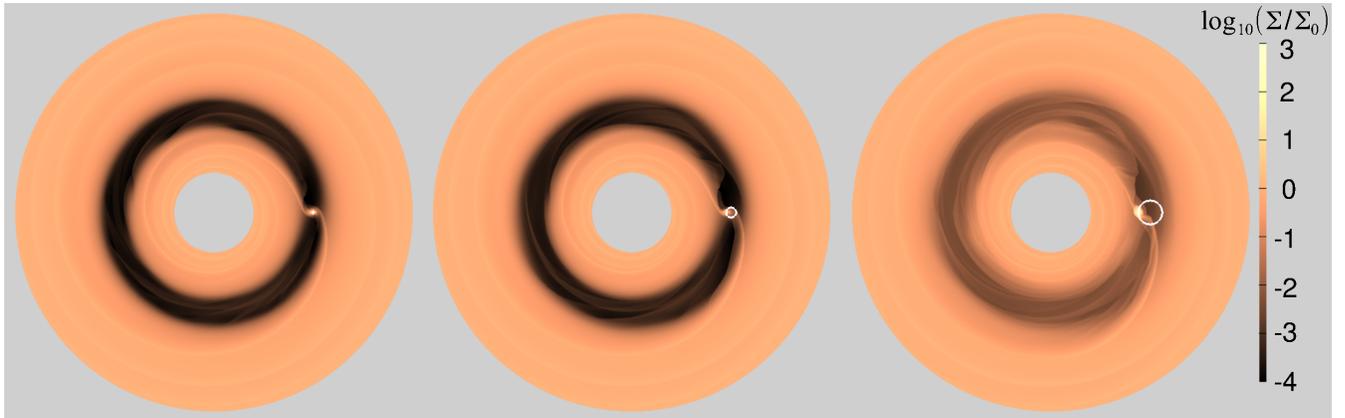}
\caption{ Standard disk-planet system employed in
this study: $q = 10^{-3}$, $\tilde \nu = \nu / (a^2
\Omega_0) = 2.5 \times 10^{-6}$, $\mathcal{M} = 28$
(equivalently, $h/a = 0.036$).
The three panels correspond to three choices of
planet eccentricity: $e = 0.01$, $0.05$, and $0.12$,
from left to right. White circles indicate the
planet's approximate epicycle.
The surface density inside the gap 
starts as low as
$\Sigma_{\rm gap}/\Sigma_0 \simeq 3 \times 10^{-4}$
at $e=0.01$ and increases with increasing $e$.
Eccentricity damps for $e=0.01$; amplifies
for $e=0.05$; and damps for $e=0.12$.
\label{fig:pretty} }
\end{figure*}

We address the problem of eccentricity evolution numerically by integrating the 2D (vertically integrated) isothermal hydrodynamic equations:

\begin{equation}
\partial_t \Sigma + \nabla \cdot (\Sigma \vec v) = 0
\end{equation}
\begin{equation}
\partial_t ( \Sigma v_j ) + \nabla \cdot ( \Sigma \vec v v_j + P \hat x_j - \nu \Sigma \vec \nabla v_j ) = - \Sigma \vec \nabla \phi
\end{equation}
\begin{equation}
P = c^2 \Sigma
\end{equation}
where $\Sigma$ is surface density, $P$ is pressure, $\vec v$ is velocity, $\nu$ is the kinematic viscosity, $c$ is the sound speed, and $\phi$ is the gravitational potential from the planet and central star.

The numerical integration is carried out using the DISCO code \citep{2011ApJS..197...15D, 2012ApJ...755....7D, 2013ApJ...769...41D}.  DISCO is a moving-mesh hydro code that is tailored for the study of disks.  Computational zones are annular wedges that shear past one another to follow the underlying flow.  By effectively subtracting off the background Keplerian flow, DISCO can provide an accurate solution for formally supersonic problems, and can integrate for long times.

The numerical domain extends from an inner radius $r_{\rm in} = 0.4$ to an outer
radius $r_{\rm out} = 2$, with the planet's semimajor axis located at radius $r = a = 1$.\footnote{Unless
otherwise indicated, we work hereafter in code
units: $G M_\ast = a = 1$ (which implies $\Omega_0 = 1$), where the variables have their usual meanings. Note that
we also set our background surface
density $\Sigma_0 = 1$, which nominally
implies a disk mass comparable to the
stellar mass; but our code ignores
self-gravity and therefore all of our
results for torques, $\dot{e}$,
and $\dot{a}$ simply scale as $\Sigma_0$.}
The domain is divided into $N_r = 360$ logarithmically spaced radial zones, corresponding
to $\Delta r / r \simeq 0.0045$. The azimuthal resolution varies with radius
to ensure grid cells with near-unity aspect ratios,
$\Delta r \simeq r \Delta \phi$.

\begin{figure*}
\epsscale{1.15}
\plotone{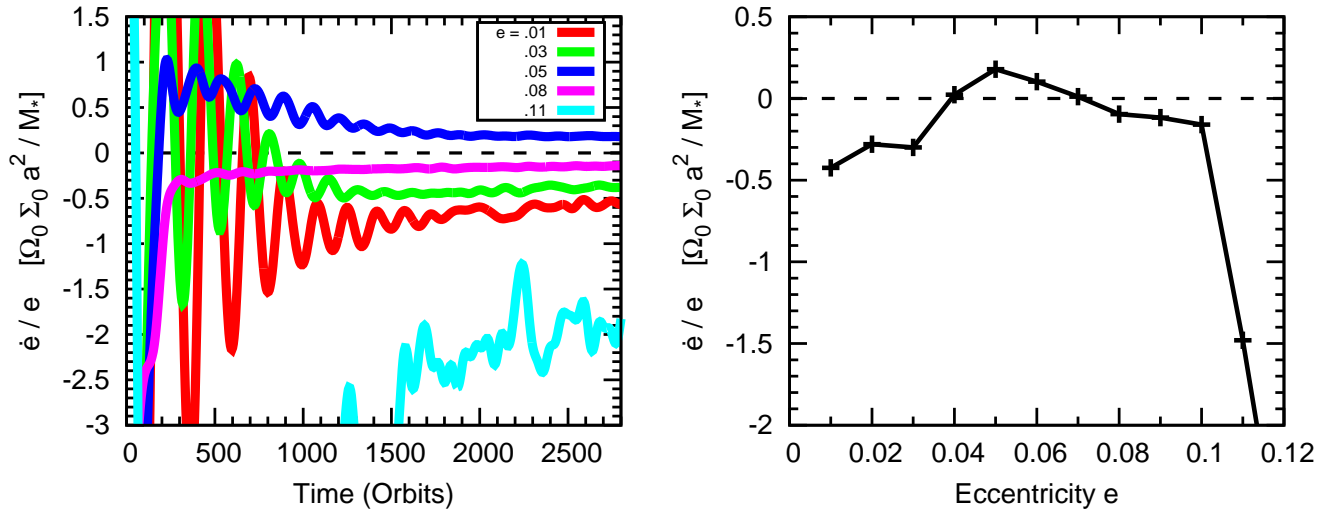}
\caption{ Time derivative of eccentricity as a
function of eccentricity for our standard
model parameters. {\it Left:}
$\dot e/e$ as a running average over time,
demonstrating convergence.
{\it Right:} Final time-averaged $\dot e/e$ as a
function of $e$. Eccentricity is damped
for $e < e_{\rm min} \simeq 0.04$; excited for
$e_{\rm min} < e < e_{\rm max} \simeq 0.07$;
and damped for $e > e_{\rm max}$. 
Thus, there are two attractors:
$e=0$ and $e=e_{\rm max}$.
Eccentricity damping
is particularly strong at $e > 0.1$ when
the planet collides into the gap walls.
\label{fig:edot} }
\end{figure*}

\subsection{Disk Model and Planet Potential}

A simple background disk is employed that ignores gradients in density $\Sigma$, viscosity
$\nu \equiv \alpha (h/a)^2$,
and sound speed $c$:

\begin{equation}
\Sigma(r) = \Sigma_0 = 1
\end{equation}
\begin{equation}
\Omega(r) = \Omega_0 (r/a)^{-3/2}
\end{equation}
\begin{equation}
v_r(r) = - \frac32 \nu / r
\end{equation}
\begin{equation}
P(r) = c^2 \Sigma_0
\end{equation}
\begin{equation}
c = a \Omega_0 / \mathcal{M}
\end{equation}
where $v_r$ is the background radial accretion velocity
and $\mathcal{M} \equiv a/h$ is the constant Mach number, with $h$ the gas scale height.

The gravitational potential at position $\vec x$ is that of the star + planet: 
\begin{equation}
\phi(\vec x) = GM_\ast \left( \frac{1}{|\vec x - \vec x_\ast|} +  \frac{q}{ \sqrt{ (\vec x - \vec x_{\rm p})^2 + \epsilon^2 }} \right)
\end{equation}
where $q = M_{\rm p}/M_\ast$ is the planet-to-star mass ratio and $\epsilon = 0.5 h$ is a smoothing length.
The positions of the planet $\vec x_{\rm p}(t)$ and star $\vec x_\ast(t)$ are found by
solving Kepler's equation for an eccentric orbit using a Newton-Raphson root-finding scheme.  Both planet and star are moved
explicitly in time, keeping the center of mass fixed at $r=0$.
Accretion onto the planet is not modeled.

Standard model parameters are $\{q,\alpha,\mathcal{M}\} = \{0.001, 0.002, 28\}$.
We also vary each of these 3 parameters separately to values above and below their standard value,
generating an extra 6 models to explore parameter space.

\subsection{Calculating $\lowercase{\dot{e}}$ Numerically}

\begin{figure}
\epsscale{1.15}
\plotone{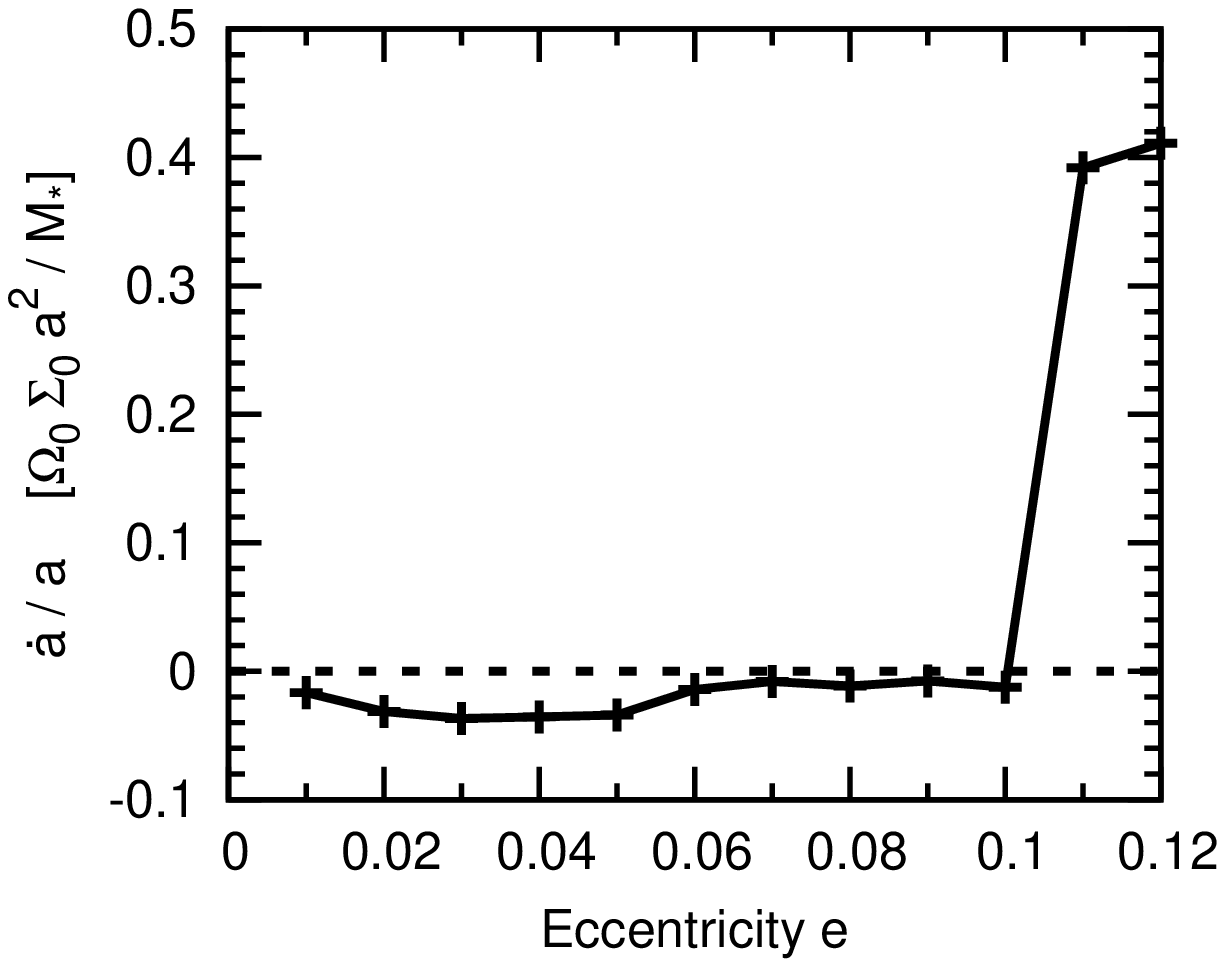}
\caption{  Migration rates as a function of
eccentricity for our standard model.  For modest
eccentricities, $\dot a$ is negative and roughly
independent of $e$.  Once the planet collides with
the gap walls, $\dot a$ becomes large and positive.
The fast outward migration coincides with the large negative
$\dot e$ seen in Figure \ref{fig:edot}.
\label{fig:adot} }
\end{figure}

\begin{figure}
\epsscale{1.15}
\plotone{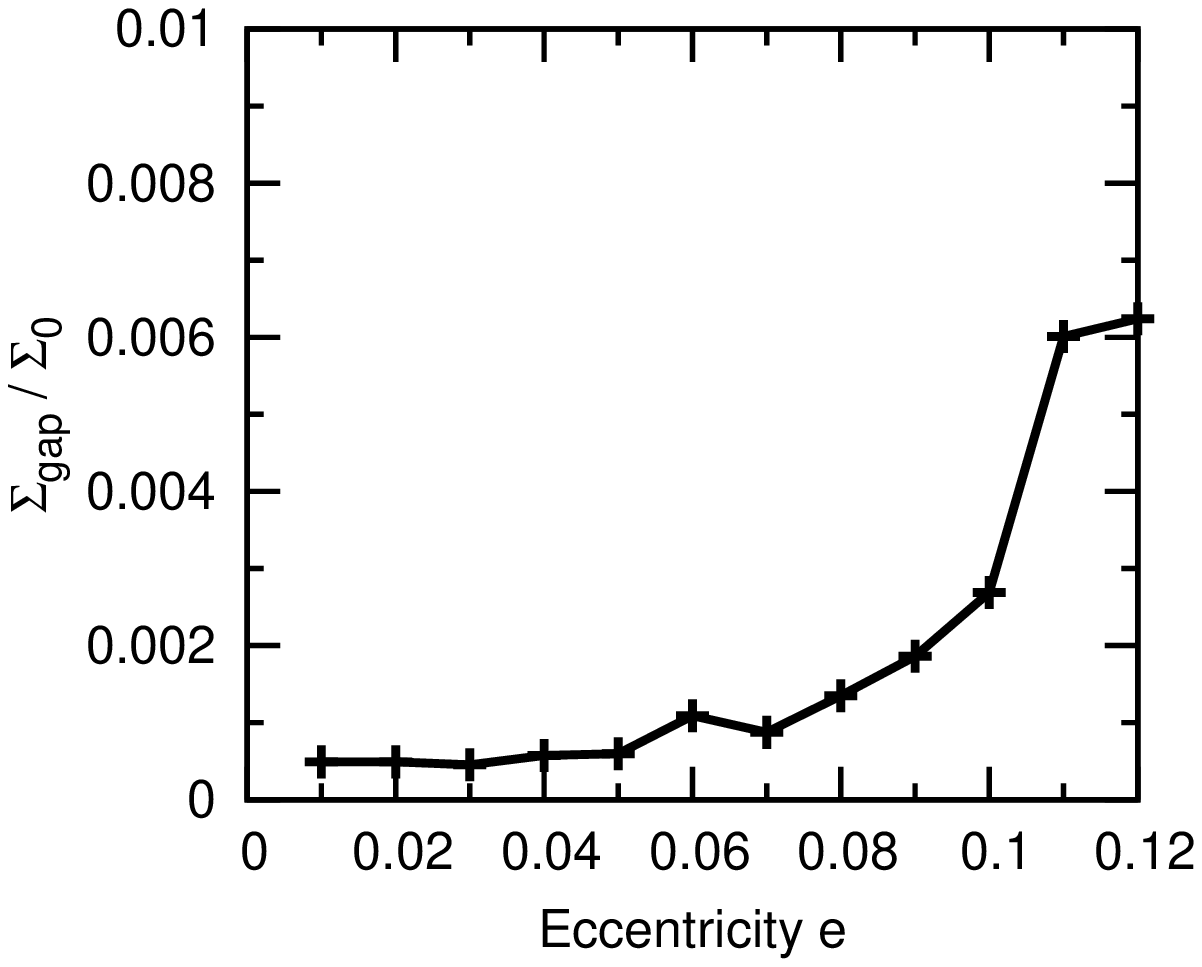}

\caption{ Gap depth $\Sigma_{\rm gap}/\Sigma_0$ is computed as a function
of eccentricity.  Depths are calculated by averaging $\Sigma(r)$ azimuthally and over time, excising from the azimuthal average a region of radius $0.2 a$ centered on the planet's guiding center.  The minimum of this averaged $\Sigma(r)$ gives $\Sigma_{\rm gap}$.
\label{fig:sigma} }
\end{figure}

The planet lives on a fixed eccentric orbit of semimajor axis $a$ and eccentricity $e$ (Figure \ref{fig:pretty}).  The code is run until the disk surface density relaxes to a
pattern that varies repeatedly and consistently with the planet's epicyclic
motion; typically this takes thousands of planetary orbits (see Figure 2). The time derivative of eccentricity is time-averaged and recorded
as a function of the chosen eccentricity, $\dot e (e)$.  

The instantaneous value of $\dot e$ follows from the definitions of the planet's
orbital angular momentum and energy:

\begin{equation}
L = a^2 \Omega_0 M_{\rm p} \sqrt{ 1 - e^2 }
\end{equation}
\begin{equation}
E = - \frac12 a^2 \Omega_0^2 M_{\rm p} \,.
\end{equation}
Combining the time derivatives of these two quantities (and remembering that $\Omega_0 = \sqrt{GM_\ast/a^3}$ depends on $a$) yields

\begin{equation}
{\dot e \over e} = { P ( 1 - e^2 ) - \Omega_0 T \sqrt{1 - e^2} \over \Omega_0^2 a^2 M_{\rm p} e^2 }
\label {eqn:edot}
\end{equation}
where $T$ and $P$ are the torque and power delivered to the planet, respectively:

\begin{equation}
T = \dot L = r F_{\theta} \label{eqn:torque}
\end{equation}
\begin{equation}
P = \dot E = \vec F \cdot \vec v_{\rm p} \,.
\end{equation}
The planet's velocity is $\vec v_{\rm p}$ and the disk's gravitational force on the planet is

\begin{equation} \label{eqn:everything}
\vec F = G M_{\rm p} \sum_{\text{zone} ~j} {   \Sigma dA_j  \over (\vec x_j - \vec x_{\rm p})^2 + \epsilon^2 } \hat l,
\end{equation}
where $\hat l$ is the unit vector pointing from the planet to the grid cell of area $dA_j$.
The planet's migration rate can also be calculated via

\begin{equation}
{\dot a \over a} = { 2 P \over \Omega_0^2 a^2 M_{\rm p} }.
\label {eqn:adot}
\end{equation}

\section{Results}
\label{sec:results}	

Results are first presented for our standard model of a Jupiter-mass planet ($q = 10^{-3}$) in a disk with $h/a = 0.036$ ($\mathcal{M} = 28$) and $\tilde \nu = \nu / (a^2 \Omega_0) = \alpha / {\mathcal M}^2 = 2.5 \times 10^{-6}$ ($\alpha = 2 \times 10^{-3}$).  Figure \ref{fig:edot} displays the time derivative of eccentricity as a function of the eccentricity, $\dot e (e)$.

The left panel shows the running time average of $\dot e / e$, demonstrating that it can take many thousands of orbits to achieve a quasi-steady state (not surprising given the low viscosity).  The right panel shows the asymptotic value of $\dot e / e$, as a function of $e$.  

Some highlights from
Figure \ref{fig:edot}:

\begin{itemize}
\item For intermediate eccentricities,
$\dot e > 0$: Jupiter-mass planets can,
under certain circumstances, have
their eccentricities excited by the disk.

\item As $e \rightarrow 0$, $\dot e < 0$.
Thus $e=0$ is an attractor of the system
for small $e$.

\item Eccentricity excitation occurs only
for $e > e_{\rm min} \simeq 0.04$: this is a
finite-amplitude instability, as predicted
by GS03.

\item As $e$ increases, eccentricity
eventually damps.
The value $e_{\rm max} \simeq 0.07$ is a
second attractor, relevant
for $e > e_{\rm min}$.

\item For the largest values of $e$
considered, $\dot{e}$ plunges to large
negative values. Here the planet's epicyclic
motion causes it to collide with the
gap walls (see also Figure
\ref{fig:pretty}); the gap fills up and
eccentricity strongly damps.


\end{itemize}

Figure \ref{fig:adot} plots the migration rate $\dot a/a$ for our standard model.  Note that in contrast with $\dot e$, the migration rate does not depend sensitively on $e$, at least until $e \gtrsim w/a \simeq 0.1$ and
the planet crashes into the gap walls.
The substantial damping of eccentricity for $e \gtrsim w/a$ found in Figure \ref{fig:edot} coincides with a large, positive migration rate in Figure \ref{fig:adot}, similar to what was
observed in the live-planet study of \cite{2006ApJ...652.1698D}.  However, it should be emphasized that this fast outward
migration is only sustained as long as the eccentricity is this large.  In reality the eccentricity should be
quickly damped to $e = e_{\rm max} \simeq 0.07$ (Figure \ref{fig:edot}), whereupon
$\dot{a} < 0$ as usual.

Figure \ref{fig:sigma} shows the gap depth
$\Sigma_{\rm gap} / \Sigma_0$ for this
system.  Gap depth is computed by calculating the azimuthally averaged and time-averaged surface density as a function of radius, $\Sigma(r)$, and finding the minimum of this function.  A region of radius $0.2a$ centered
on the planet's guiding center at 
$(r,\theta) = (a,\Omega_0 t)$ is excised from the azimuthal average, in order to avoid contamination from material very close to the planet.  Increasingly eccentric planets have shallower gaps.

In the next section, we elaborate upon all
the trends highlighted above. We apply
the theory of disk-planet interactions
pioneered by \cite{1980ApJ...241..425G}
to see if we can reproduce quantitatively
the behavior of $\dot{e}$ measured
numerically.

\subsection{Detailed Comparison with GS03 \\for Standard Model}
\label{sec:anly}

Here we compare our numerical results for $\dot{e}$ for our standard model with those from analytic theory.  Using formulae derived by \cite{1980ApJ...241..425G}, \cite{1988Icar...73..330W}, \cite{2000MNRAS.315..823P}, and \cite{2003ApJ...587..398O},
we compute the contributions
to $\dot{e}$ from various kinds of resonances:
principal Lindblad resonances,
first-order (as expanded in the planet's eccentricity)
Lindblad resonances, and first-order co-rotation resonances.
Principal co-rotation resonances are omitted from our analysis,
as these depend on $d\Sigma/dr$ at the very gap center;
this derivative (difficult to calculate reliably) 
is assumed to be negligibly small for our deep gaps.

\begin{figure}
\epsscale{1.15}
\plotone{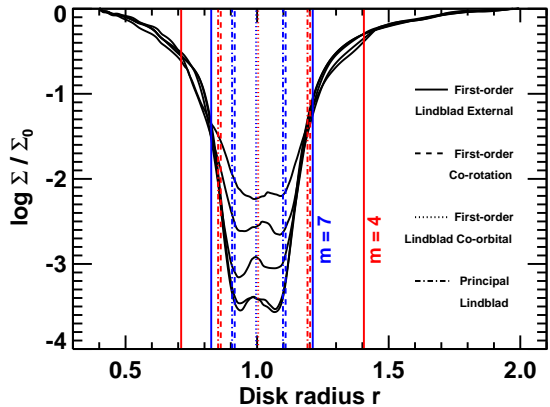}
\caption{How gap profiles vary with $e$ for our standard model. From bottom to top, surface density
profiles correspond to $e = 0.01$, 0.03, 0.06, 0.10, and 0.12.
Each profile is azimuthally averaged from a late-time 
snapshot excised of a circular region of radius = $0.2a$
centered on $(r,\theta) = (a,\Omega_0 t)$; the excision removes
the highly overdense material in the planet's immediate vicinity
from the azimuthal average. Resonances from two of the more
significant wavenumbers (as judged from Figure \ref{fig:cumu})
are plotted. What appear to be nearly overlapping resonances in the figure actually do completely
overlap in their nominal positions (e.g., the $m=4$ principal
Lindblad and $m=4$ first-order co-rotation resonances);
we plot these overlapping resonances with small arbitrary offsets for visual
clarity only. The co-orbital resonances are so named because
they are located at the planet's semi-major axis ($r=a$);
they should not be confused with the
co-rotation resonances, which are offset 
from $r=a$ because they co-rotate with a particular term in the planet's
Fourier-expanded potential whose pattern speed does not in general
equal the planet's mean motion.
\label{fig:sigma_prof} }
\end{figure}

\begin{figure}
\epsscale{1.15}
\plotone{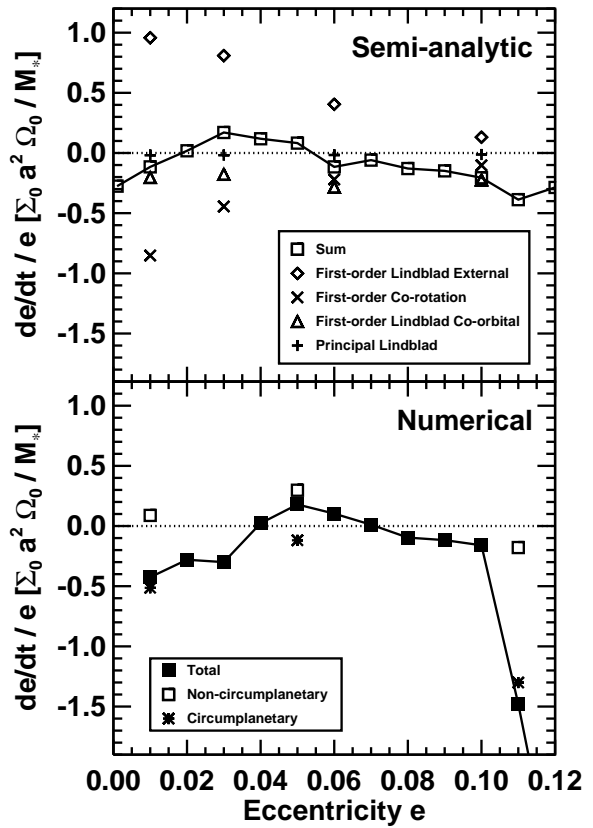}
\caption{Disentangling the web of resonances that contribute
to eccentricity evolution for our standard disk model.
{\it Top}: $\dot{e}/e$ as computed semi-analytically
from the formulae in the Appendix,
evaluated using the azimuthally
averaged surface density profiles $\Sigma(r)$ from our numerical calculations (Figure \ref{fig:sigma_prof}). For $e=0.01$, we employ
$\Sigma(r)$ as computed for $e=0.1$, since
the surface density profiles do not change much
for $e \leq 0.03$. {\it Bottom}: $\dot{e}/e$
computed wholly numerically, with contributions
from circumplanetary (within 1 Hill radius) and non-circumplanetary
material distinguished for a few sample $e$'s. 
The semi-analytic calculation exhibits two trends: (1) a rise
in $\dot{e}/e$ at small $e$ accompanied by a zero crossing
that reflects the saturation of first-order co-rotation resonances
and the growing dominance of first-order Lindblad
external resonances; and (2) a drop in $\dot{e}/e$
at large $e$ accompanied by a second zero crossing that
reflects the weakening of Lindblad resonances from the planet's increasingly
supersonic epicyclic motion. These behaviors appear qualitatively
reproduced by the numerical calculations, with modifications introduced
by circumplanetary torques that linear theory does not capture. The huge drop in $\dot{e}$ at $e \geq 0.11$
arises from the planet careening into the gap walls.
\label{fig:semi} }
\end{figure}

\begin{figure}
\epsscale{1.15}
\plotone{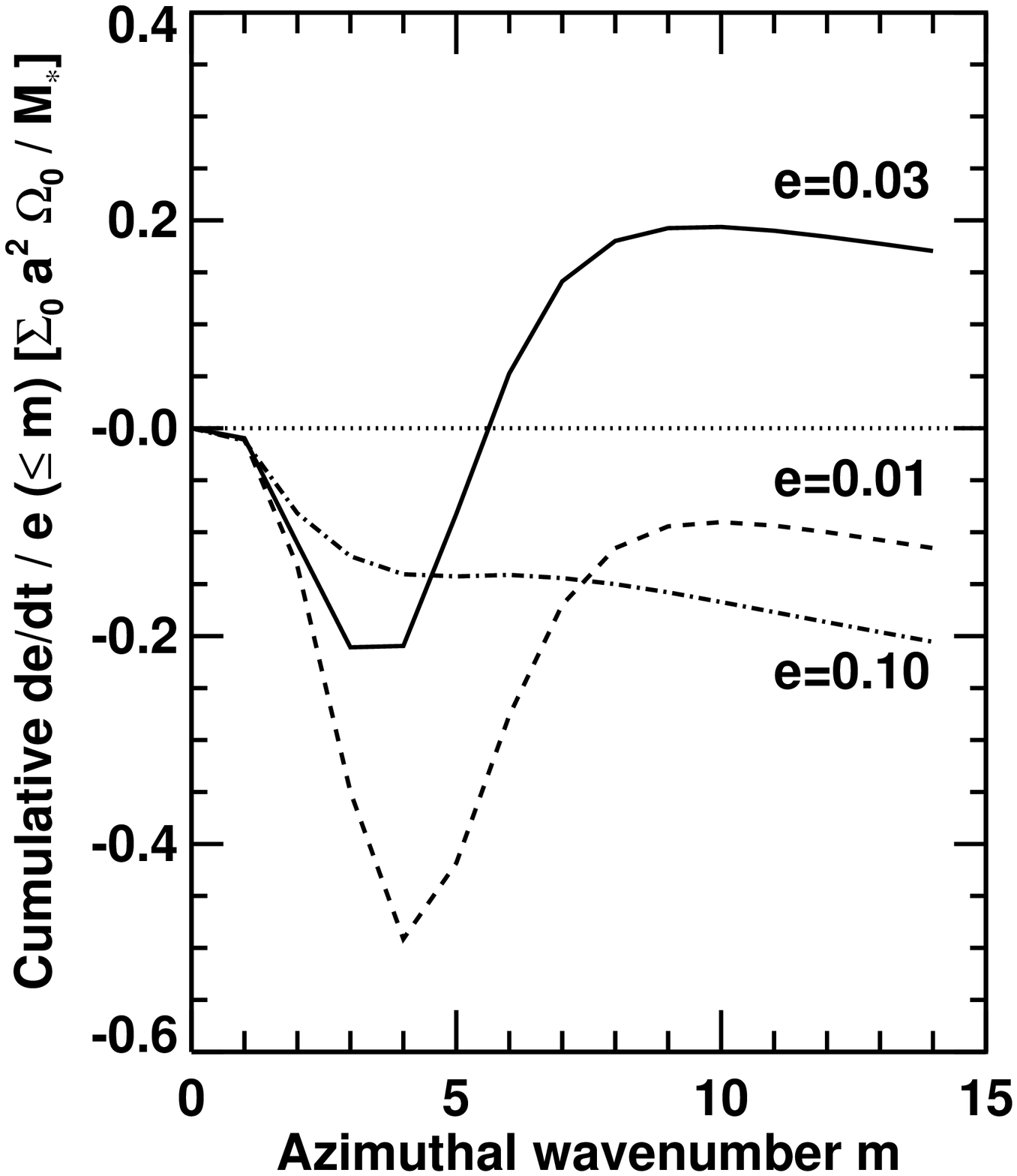}
\caption{Running sum of $\dot{e}/e$ vs.~azimuthal
wavenumber $m$ for our standard disk model 
parameters, calculated semi-analytically.
Contributions from all kinds of resonances (see
Figure \ref{fig:semi})
are totaled for every $m$. The sum is truncated at
$m_{\rm max} = \mathcal{M}/2$ = 14 to crudely
account for the ``torque cut-off" \citep{1980ApJ...241..425G}.
Most of the contributions to $\dot{e}/e$ arise from
$m \simeq 2$--8.
The locations of the various resonances for $m=4$
and $m=7$ are shown
in Figure \ref{fig:sigma_prof}.
\label{fig:cumu} }
\end{figure}

The formulae for $\dot{e}$ are given in the Appendix.
They depend on surface density $\Sigma (r)$ and its slope
$d\Sigma(r)/dr$;
these two quantities are read directly off snapshots of the numerical solution,
so in this sense our calculation is semi-analytic.\footnote{In the
case of a gapless disk
($\Sigma = \Sigma_0$), the
equations in the Appendix
give a value for $\dot{e}/e < 0$ that
matches that of equation (\ref{eqn:edote})
to within $\sim$20\%, after adjusting
the strength of the softening term in the generalized
Laplace coefficient.}
A sampling of surface density profiles $\Sigma(r)$ vs.~$e$ 
is provided in Figure \ref{fig:sigma_prof}, 
overlaid with the locations of some of the more
important resonances. Each surface density
profile is taken from an individual
snapshot in time, azimuthally averaged after excising
a circular region of radius = $0.2a$ centered on the planet's guiding center at $(r,\theta) = (a,\Omega_0 t)$.
The excised region contains large and highly time-variable
overdensities in the immediate vicinity of the planet
that the analytic theories---which govern small disturbances on a
smooth background---were not intended to treat.
We will see at the end of this section that torques from this excised region are 
significant in some regions
of parameter space.

The contributions to $\dot{e}$ from the various
resonances are dissected in Figure \ref{fig:semi} (top panel).
As anticipated by GS03,
the strongest resonances are the
first-order Lindblad external resonances which excite $e$,
and the first-order co-rotation resonances which damp $e$.
The first-order Lindblad co-orbital resonances
also damp $e$, but are weaker because they are situated
in the dead center of the gap where surface densities
are at their lowest. Principal Lindblad resonances contribute negligibly to $\dot{e}$.
Figure \ref{fig:cumu} shows that wavenumbers $m \simeq 2$--8
contribute most to $\dot{e}$; contributions from higher $m$,
up to our assumed cut-off at $m_{\rm max} = \mathcal{M}/2$,
are less important.

The broad similarity between our semi-analytic calculation (Figure
\ref{fig:semi}, top panel) and our numerical results
(Figure \ref{fig:semi}, bottom panel) emboldens us to
give the following interpretation of the dynamics.
As $e$ increases from 0, $\dot{e}$ switches from negative
to positive. This first zero crossing is the finite-amplitude instability of GS03 and \cite{2003ApJ...587..398O}.
The instability results because the first-order
co-rotation resonances (which damp $e$) weaken from increasing saturation
with increasing $e$ --- i.e., they weaken following
the $F(p)$ saturation function 
(see Appendix equations \ref{preog03}--\ref{og03}, plus the
discussion at the end of this subsection).
Above a threshold $e$, the
co-rotation resonances give
way to the first-order Lindblad external resonances which render
$\dot{e}$ positive in the net. Further increases
in $e$, however, bring $\dot{e}$ back down to a second zero crossing.
The external resonances weaken
as $e$ exceeds $h/a$, i.e., as the planet's epicyclic
motion becomes supersonic \citep{2000MNRAS.315..823P}.
The consequence of this supersonic weakening with
increasing $e$ is that 
the first-order co-rotation resonances---which are unaffected by
supersonic motion---together with the first-order Lindblad co-orbital resonances 
regain the upper hand at large $e$ to make $\dot{e} < 0$.
Although the co-orbital
Lindblads suffer from the same supersonic
weakening as do the external Lindblads, the co-orbitals
yield a (negative) value of
$\dot{e}/e$ that hardly varies with $e$;
their weakening is mitigated by the surface density
at gap center which grows with $e$ (Figures \ref{fig:sigma} and \ref{fig:sigma_prof}),
maintaining the strength of the co-orbitals.

Perhaps the most glaring discrepancy between our semi-analytic and numerical results
is at the largest values of $e$. Numerically, at $e\geq 0.11$,
we find eccentricity damping rates that are substantially higher
than those expected from theory. As the bottom panel of
Figure \ref{fig:semi} indicates,
the large negative values of $\dot{e}/e$ are generated 
from torques exerted by ``circumplanetary" material---here defined as material within 1 Hill radius of the planet's instantaneous position (the circumplanetary region so defined
is a subset of the excised region used to calculate
azimuthal averages of surface density).
The dominance of circumplanetary torques is not surprising
at large $e$: the epicyclic motion is so wide that the planet
collides with the gap walls and suffers dynamical friction from high density gas.
What is surprising is that circumplanetary torques also dominate at the smallest
value of $e=0.01$, rendering $\dot{e}$ more negative than expected from theory
and pushing the onset of the eccentricity instability to larger
$e \simeq 0.04$ (Figure \ref{fig:semi}, bottom panel).
The properties of the circumplanetary region are uncertain and cannot
be reliably predicted from linear theory.
In our numerical calculations, details of the circumplanetary
flow are subject to such
issues as grid resolution, smoothing length,
and prescriptions
for how the planet accretes.

Returning to the first zero-crossing for $\dot{e}$
as computed semi-analytically (top panel
of Figure \ref{fig:semi}), we reiterate
that it occurs because of co-rotation
saturation, as quantified by $F(p)$ (equation \ref{og03}).
This saturation function from \cite{2003ApJ...587..398O}
(called $t_c(p)$ by them) decreases with increasing $e$;
it causes the $\times$ symbols in the top panel
of Figure \ref{fig:semi} to approach zero
as $e$ increases. Physically, $F(p)$
describes how the co-rotation torque
weakens as viscous diffusion is increasingly unable
to supply the co-rotation region with fresh
librating material. The saturation function
is separate from the surface density gradient
$d\Sigma/dr$ (really vortensity gradient)
which also factors into the strength
of the co-rotation torque (equation \ref{dsdr}).
The surface density gradient at the location
of a co-rotation resonance also decreases
as $e$ increases (see Figure \ref{fig:sigma_prof}),
but the decrease in $d\Sigma/dr$, in and of itself,
is not as significant as the decrease in $F(p)$.
We have shown this by re-computing $\dot{e}/e$ vs.~$e$
using the single surface density profile $\Sigma(r)$
evaluated for $e=0.01$, and obtaining a curve
similar to the one shown in the top panel of Figure
\ref{fig:semi}.

\subsection{Dependence on Disk and Planet Parameters}
\label{sec:param}

\begin{figure}
\epsscale{1.15}
\plotone{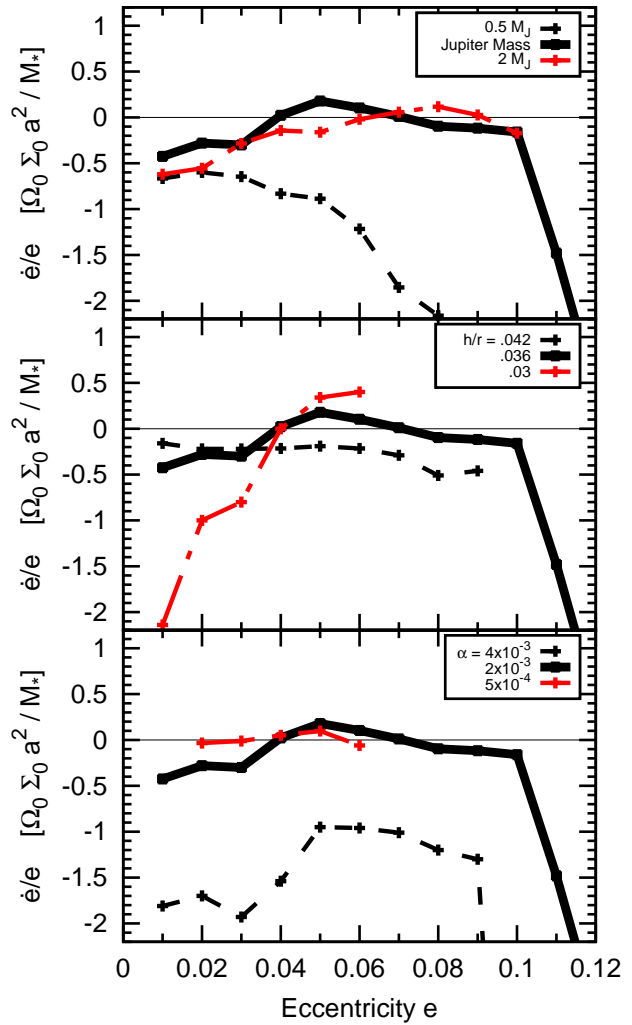}
\caption{How $\dot{e}/e$ varies across parameter
space. Data are plotted only for those models
whose running time-averages of $\dot{e}/e$
converged to well-defined values; disks with
especially low $h/a$ or low $\alpha$ perturbed
by planets with high eccentricity exhibited
strong instabilities and failed to give
convergent answers. Eccentricity driving
favors large planet masses, small $h/a$,
and small $\alpha$, the same region of
parameter space that produces deep gaps.
\label{fig:param} }
\end{figure}

\begin{figure}
\epsscale{1.15}
\plotone{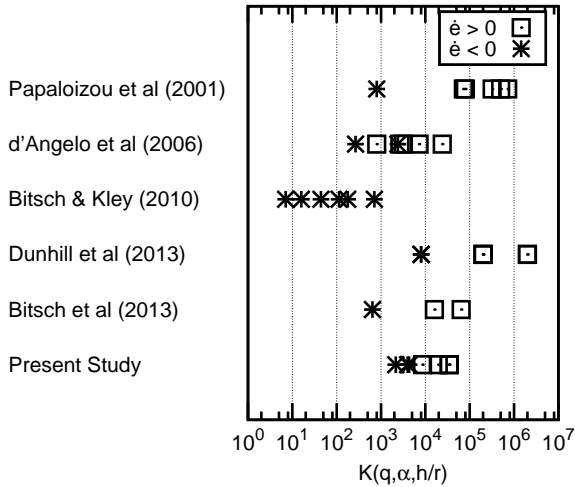}
\caption{ Several studies of planet-disk
eccentricity evolution are compared using the
parameter $K \equiv q^2 \mathcal{M}^5 / \alpha$,
which governs gap depths \citep{2013ApJ...769...41D, 2014ApJ...782...88F}.
A rough correlation between high
$K \gtrsim 10^3$--$10^4$ (deep gaps having
$\Sigma_{\rm gap}/\Sigma_0 \lesssim 10^{-3}$) and
$\dot{e}>0$ can be discerned.
\label{fig:K} }
\end{figure}

Figure \ref{fig:param} shows results for
$\dot{e}(e)$ for our six disk-planet systems
scattered across parameter space. The
observed changes to
$\dot{e}(e)$ are complicated and difficult
to follow in detail. We performed
the same semi-analytic analysis
for these models as we did for our standard
model (Section \ref{sec:anly}),
and were able to reproduce
only some of the trends documented
in Figure \ref{fig:param}. Part of our
difficulty stemmed from circumplanetary
torques which often proved significant,
and which we could not evaluate
using the standard analytic theory.

Broadly speaking, however,
we can say that $\dot{e} > 0$ favors
high-mass planets, thin disks,
and low viscosities --- these cases
are highlighted in red
in Figure \ref{fig:param}.

Qualitatively, the circumstances that
lead to $\dot{e}>0$ are the same ones that
produce deep gaps.
Gap depths are gauged by the
dimensionless parameter
\begin{equation}
K(q,\alpha,h/a) = { q^2 \over \alpha (h/a)^5 }.
\end{equation}
\citep{2013ApJ...769...41D,
2014ApJ...782...88F, 2015MNRAS.448..994K,
2015ApJ...807L..11D} and it is interesting
to ask whether this same parameter
can predict eccentricity growth.
Figure \ref{fig:K} lists values of $K$ for our seven disk-planet parameter studies,
together with $K$-values from
previous studies of eccentricity evolution.
A rough threshold of $K \sim 10^3$--$10^4$
dividing $\dot{e}<0$ from $\dot{e}>0$
can be discerned --- equivalent to
a threshold gap depth $\Sigma_{\rm gap}/\Sigma_0 \sim 10^{-3}$.

\subsection{Gap Widths}

\begin{figure}
\epsscale{1.0}
\plotone{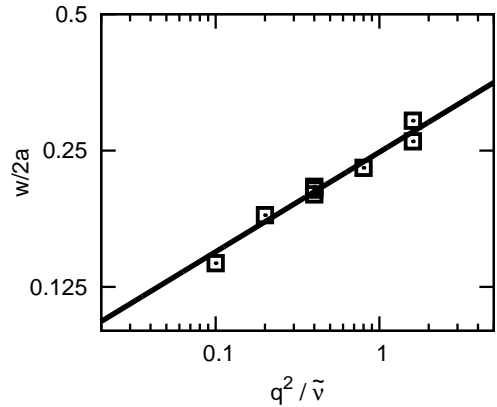}
\caption{ Gap half-widths $w/2a$
(defined as half the
distance between points in the gap where
$\Sigma/\Sigma_0$ = 0.1) increase with increasing
planet-to-star mass ratio $q$ and decreasing
disk viscosity $\tilde \nu$. Equation
(\ref{eq:width}) predicts that
$w/2a \sim (q^2/\tilde \nu)^{1/3}$;
our measurements are fitted by 
$w/2a \simeq 0.25 (q^2/\tilde \nu)^{0.22}$
(solid line). The gap half-width gives a hard
upper bound on eccentricities
that can be sustained by planets
embedded in their natal gas disks.
The actual bound $e_{\rm max}$
on $e$ is somewhat smaller ---
of order a few times $h/a$ --- and occurs
when supersonically-weakened external resonances
exactly cancel co-orbital and co-rotation resonances
to render $\dot{e}=0$.
\label{fig:width} }
\end{figure}

The width of a gap opened by a planet
gives a hard upper bound on the
eccentricity that can be excited
by disk torques. As such, it is worthwhile
understanding how the gap width $w$
depends on input parameters.

The exercise performed in Section 2
predicts that
$w/a \sim (q^2/\tilde \nu)^{1/3}$.
This relation is tested in Figure
\ref{fig:width} where gap half-widths
are plotted against
the dimensionless parameter
$q^2 / \tilde \nu$.
The gap half-width is 
evaluated by differencing
the radii at which $\Sigma = 0.1 \Sigma_0$ and dividing by two. The data in Figure
\ref{fig:width}
appear to conform to a power law,
but with a somewhat shallower slope
than the predicted 1/3:
$w/2a = 0.25 (q^2/\tilde\nu)^{0.22}$,
where $w/2$ is the gap half-width.

For our standard model parameters,
the above fitting formula gives
$w/2a = 0.2$.
By comparison, the eccentricity
beyond which $\dot{e}$
plummets to large negative values
(see Figure \ref{fig:edot}) is $e = 0.11$;
this is a factor of 2 smaller
than $w/2a$ and
suggests that in this context,
a more relevant definition
for gap half-width 
would be obtained by taking
$\Sigma/\Sigma_0 = 10^{-3}$ rather
than $\Sigma/\Sigma_0 = 10^{-1}$ --- see Figure \ref{fig:sigma_prof}.

We close with the reminder that the actual
value to which a planet's eccentricity
relaxes is not given by the gap-collision
value, but rather by
$e_{\rm max}$ (which for our standard
model equals 0.07) --- this is the
value for which $\dot{e} = 0$, and marks
where supersonically weakened external
Lindblad resonances balance
co-rotation and co-orbital Lindblad
resonances (Section \ref{sec:anly}).

\section{Summary and Discussion}
\label{sec:discussion}


This study demonstrates that
Jupiter-mass planets can have their
orbital eccentricities amplified by disk
torques --- provided they open deep
enough gaps, and provided their
eccentricities exceed a threshold value.
The finite-amplitude instability documented
here appears to be
the same as that predicted
analytically by \cite{2003ApJ...585.1024G}.
Eccentricities are damped
by first-order co-orbital Lindblad
torques and first-order co-rotation
torques. Deep gaps are required to
disable the former, while
finite eccentricities serve
to saturate (i.e., weaken)
the latter \citep{2003ApJ...587..398O}.
With these requirements met,
first-order external Lindblad resonances
can excite a planet's eccentricity.

Our results are similar to those of \cite{2006ApJ...652.1698D} who also found eccentricity growth for Jovian-mass planets
at low disk viscosities ($\alpha \sim 10^{-3}$), but differ from them insofar
as the eccentricity growth that
we report explicitly
requires a non-zero initial
eccentricity.
Other studies did not
find eccentricity growth for
Jupiter-mass planets but used
larger viscosities or thicker disks.
Their results may be reconciled
with ours by examining the parameter
$K \equiv q^2 \mathcal{M}^5/\alpha$
which governs gap depth.
Eccentricity driving seems to require
large $K \gtrsim 10^3$--$10^4$,
i.e., deep gaps of surface
density
$\Sigma_{\rm gap}/\Sigma_0 \sim 1/K$.

Eccentricities do not amplify
without bound. As eccentricities
increase above the disk aspect ratio
$h/a$ --- i.e., as the planet's epicyclic motion
becomes supersonic --- the external
resonances weaken
\citep{2000MNRAS.315..823P}. At the
same time, the co-orbital resonances
strengthen with increasing eccentricity
as more disk material
leaks into the gap. Consequently,
eccentricity damps above a certain
value that scales like $h/a$; this value is an attractor
for the system. In our numerical
experiments, the attractor eccentricity
ranges from 0.07 to 0.09. At still larger
eccentricities --- so large that
the planet collides into gap walls
separated by a fractional width $w/a$ --- the
damping of eccentricity
becomes catastrophically rapid.

The results of our numerical study
align with analytic expectations only
broadly. Significant torques
are exerted by material within a Hill
sphere or so of the planet that
linear theory cannot capture.
Modeling circumplanetary flows
is technically challenging and we
do not claim to have gotten it right.
In addition to the usual worries about
smoothing lengths and planetary accretion
prescriptions, there looms the possibility
that flows in 3D could look
qualitatively different from our 2D
solutions \citep{2015MNRAS.447.3512O}.
In particular, gaps may be systematically
shallower and co-rotation resonances might never
saturate \citep{2015arXiv150503152F}.

The eccentricity driving reported
here does not rely on
non-barotropic (i.e., non-isothermal or non-adiabatic) thermodynamics,
as our numerical calculations are
for strictly isothermal disks.
Including non-barotropic effects
such as those introduced by external
irradiation of gap walls may help
to lower the threshold eccentricity
for instability \citep{2014ApJ...782..113T}.
If eccentricity driving survives
in 3D, it offers a possible
explanation for the low-to-moderate
eccentricities $\lesssim 0.1$
observed for giant planets---including 
Jupiter and conceivably Saturn---without
recourse to planet-planet
interactions \citep[cf.][]{2005Natur.435..459T}.

\acknowledgments

Computational resources
were provided by NASA's High-End Computing
(HEC) Program administered by the
NASA Advanced Supercomputing (NAS) Division
at Ames Research Center. Financial support
was provided by the Theoretical Astrophysics
Center and the Center for Integrative
Planetary Science at UC Berkeley, and by
grants to EC from the NASA Origins and NASA
Outer Planets programs. We are grateful to
an anonymous referee for an encouraging report,
and Bertram Bitsch,
Rebekah Dawson, Alex Dunhill, Steve Lubow,
Alessandro Morbidelli,
Re'em Sari, and Taku Takeuchi for helpful discussions.

\bibliographystyle{apj}


\appendix

\section{Semi-Analytic Calculation of $\lowercase{\dot{e}/e}$}

We list here the formulae used to evaluate
$\dot{e}$ and make Figures
\ref{fig:semi} and \ref{fig:cumu}.
We let $a_{\rm planet}$ be the planet's semi-major
axis (this is the same variable as $a$ in the main
text), $e$ the planet's eccentricity,
$M_\ast$ the stellar mass,
and $G$ the gravitational constant.
We further define
$H_{\rm planet} = M_{\rm planet} \sqrt{G M_\ast a_{\rm planet} (1 - e^2)}$ to be the planet's orbital
angular momentum and $\Omega_{\rm planet} = \sqrt{GM_\ast/a_{\rm planet}^3}$
to be the planet's orbital angular frequency.
In all our numerical evaluations, $G M_\ast=a_{\rm planet}=\Omega_{\rm planet}=1$. The disk
semi-major axis is $r$.

We account for eight kinds of resonances: all
those considered in Table 1
of \cite{2003ApJ...585.1024G} with the exception
of the principal co-rotation resonance.
For every resonance type, we give the full set of
equations required to compute $\dot{e}$.
Some of the equations are shared between
types, but we list the complete set anyway under
each type for ease of reference.

\subsection{First-order Lindblad resonances\\with $\ell=m+1$}
The pattern speed of the potential
component with $\ell = m+1$ where $m$
is the azimuthal wavenumber (see \cite{1980ApJ...241..425G} for their Fourier notation) is given
by
\begin{equation}
\Omega_{m+1,m} = \left( \frac{m+1}{m} \right) \Omega_{\rm planet}\,.
\end{equation}
There are two Lindblad resonances associated with this
potential: a ``co-orbital" resonance so called because
it is located at the planet's semi-major axis:
\begin{equation}
\begin{rcases}
\beta \equiv r/a_{\rm planet} = 1 \\
\Omega = \Omega_{\rm planet}
\end{rcases}
\text{\,\,co-orbital}
\end{equation}
and an ``external" resonance:
\begin{equation}
\begin{rcases}
\beta = \left( \frac{m-1}{m+1} \right)^{2/3} \\
\Omega = \left( \frac{m+1}{m-1} \right) \Omega_{\rm planet}
\end{rcases}
\text{\,\,external.}
\end{equation}
We make no accounting for changes
in resonance location from disk pressure gradients
or self-gravity.

The planet's eccentricity changes at the rate
\begin{align} \label{eq:edot}
\dot{e} & = -\frac{\Omega_{\rm planet}}{m} \frac{H_{\rm planet}^2}{(GM_\ast)^2} \frac{T^{\rm L}_{m+1,m}}{e M_{\rm planet}^3} \times \frac{1}{1 + 0.25 \left(e \mathcal{M} \right)^3} \,.
\end{align}
The second factor involving the Mach number $\mathcal{M} \equiv a_{\rm planet} / h$ is taken from \cite{2000MNRAS.315..823P} (their equation
32), and accounts for how Lindblad torques weaken
as the planet's epicyclic motion becomes increasingly
supersonic. The first-order Lindblad torque equals
\begin{align}
T^{\rm L}_{m+1,m} & = -m \pi^2 \Sigma \left(r \frac{dD}{dr} \right)^{-1} \times \nonumber \\
& \left(r \frac{d\phi_{m+1,m}}{dr} + \frac{2\Omega}{\Omega-\Omega_{m+1,m}}\phi_{m+1,m} \right)^2 
\end{align}
where
\begin{equation}
r \frac{dD}{dr} = -3\Omega^2 + 3\Omega m^2 \left(\Omega - \frac{m+1}{m} \,\Omega_{\rm planet} \right) \,.
\end{equation}
All quantities are evaluated at the resonance location
(either equation A2 or A3). The surface density
$\Sigma$ is interpolated
from the azimuthally averaged density
profile of an excised snapshot (see Figure \ref{fig:sigma_prof} and related
text for details). The potential amplitude is given by
\begin{align}
\phi_{m+1,m} & = -\frac{GM_{\rm planet}e}{a_{\rm planet}} \times \nonumber \\
& \left[ \left(\frac{1}{2} + m + \frac{\beta}{2} \frac{d}{d\beta} \right) b_{1/2}^m(\beta) -2\beta\delta_{m,1} \right]
\end{align}
where $\delta_{m,1}$ is the Kronecker delta.
Note that $r \, d/dr = \beta \, d/d\beta$.
The Laplace coefficient is 
\begin{equation} \label{eq:laplace}
b_{1/2}^m (\beta) = \frac{2}{\pi} \int_0^\pi \frac{\cos m\phi}{(1-2\beta \cos\phi + \beta^2 + 2/\mathcal{M}^{2})^{1/2}} d\phi
\end{equation}
and is evaluated numerically.
The factor of $2/\mathcal{M}^2$ accounts roughly
for how the planet's point-mass potential is softened
by length $\sim$$h$ \citep[cf.][]{1988Icar...73..330W}.
The coefficient of $2$ is obtained by
calibrating our final answer for $\dot{e}/e$
for the test case of a gapless ($\Sigma = \Sigma_0$)
disk to match approximately the value given by
equation (\ref{eqn:edote});
adopting a smaller coefficient would
overestimate the strength of the co-orbital
resonances and lead to excessively negative
values of $\dot{e}$.

Equation \ref{eq:edot} is summed from
$m_{\rm min}=1$ to $m_{\rm max}=\mathcal{M}/2$,
where $m_{\rm max}$ crudely approximates the torque
cut-off \citep{1980ApJ...241..425G}.
For $m=1$ the external resonance
does not exist.

\subsection{First-order Lindblad resonances\\with $\ell=m-1$}
This case is analogous to the one above. The sum
runs from $m_{\rm min}=2$ to $m_{\rm max} = \mathcal{M}/2$.

\begin{equation}
\Omega_{m-1,m} = \left( \frac{m-1}{m} \right) \Omega_{\rm planet}
\end{equation}

\begin{equation}
\begin{rcases}
\beta = 1 \\
\Omega = \Omega_{\rm planet}
\end{rcases}
\text{\,\,co-orbital}
\end{equation}

\begin{equation}
\begin{rcases}
\beta = \left( \frac{m+1}{m-1} \right)^{2/3} \\
\Omega = \left( \frac{m-1}{m+1} \right) \Omega_{\rm planet}
\end{rcases}
\text{\,\,external}
\end{equation}

\begin{align} 
\dot{e} & = +\frac{\Omega_{\rm planet}}{m} \frac{H_{\rm planet}^2}{(GM_\ast)^2} \frac{T^{\rm L}_{m-1,m}}{e M_{\rm planet}^3} \times \frac{1}{1 + 0.25 \left(e \mathcal{M} \right)^3}
\end{align}

\begin{align}
T^{\rm L}_{m-1,m} & = -m \pi^2 \Sigma \left(r \frac{dD}{dr} \right)^{-1} \times \nonumber \\
& \left(r \frac{d\phi_{m-1,m}}{dr} + \frac{2\Omega}{\Omega-\Omega_{m-1,m}}\phi_{m-1,m} \right)^2 
\end{align}

\begin{equation}
r \frac{dD}{dr} = -3\Omega^2 + 3\Omega m^2 \left(\Omega - \frac{m-1}{m} \,\Omega_{\rm planet} \right) 
\end{equation}

\begin{align}
\phi_{m-1,m} & = -\frac{GM_{\rm planet}e}{a_{\rm planet}} \times \nonumber \\
& \left(\frac{1}{2} - m + \frac{\beta}{2} \frac{d}{d\beta} \right) b_{1/2}^m(\beta)
\end{align}

\noindent where $b_{1/2}^m(\beta)$ is given
by \ref{eq:laplace}.

\subsection{First-order Co-rotation Resonances\\with $\ell=m+1$}
The sum runs from $m_{\rm min}=1$
to $m_{\rm max} = \mathcal{M}/2$.

\begin{equation}
\Omega_{m+1,m} = \left( \frac{m+1}{m} \right) \Omega_{\rm planet}
\end{equation}

\begin{equation}
\begin{rcases}
\beta = \left( \frac{m}{m+1} \right)^{2/3} \\
\Omega = \left( \frac{m+1}{m} \right) \Omega_{\rm planet}
\end{rcases}
\text{\,\,co-rotation}
\end{equation}

\begin{align} 
\dot{e} & = -\frac{\Omega_{\rm planet}}{m} \frac{H_{\rm planet}^2}{(GM_\ast)^2} \frac{T^{\rm C}_{m+1,m}}{e M_{\rm planet}^3} \times F(p) \label{preog03}
\end{align}

\noindent The saturation factor is derived
by \cite{2003ApJ...587..398O} and fitted
numerically by \cite{2003ApJ...585.1024G}:
\begin{equation} 
F(p) = \frac{(1 + 0.65 p^3)^{5/6}}{(1 + 1.022p^2)^2}
\end{equation}

\begin{equation}
p = \frac{2\phi_{m+1,m}}{3\Omega^2} \left( \frac{3\Omega m}{2\beta a_{\rm planet}\nu} \right)^{2/3}
\end{equation}

\begin{align}
T^{\rm C}_{m+1,m} & = -\frac{4\pi^2 m}{3} \left( \frac{m}{m+1} \right)^2 \beta a_{\rm planet} \Omega_{\rm planet}^2 \times \nonumber \\
& \phi_{m+1,m}^2 \left( \frac{d\Sigma}{dr} + \frac{3}{2} \frac{\Sigma}{r} \right) \label{dsdr}
\end{align}

\begin{align}
\phi_{m+1,m} & = -\frac{GM_{\rm planet}e}{a_{\rm planet}} \times \nonumber \\
& \left[ \left(\frac{1}{2} + m + \frac{\beta}{2} \frac{d}{d\beta} \right) b_{1/2}^m(\beta) -2\beta\delta_{m,1} \right] \label{og03}
\end{align}

\noindent where $b_{1/2}^m(\beta)$ is given
by \ref{eq:laplace}.

\subsection{First-order Co-rotation Resonances\\with $\ell=m-1$}
The sum runs from $m_{\rm min}=2$
to $m_{\rm max} = \mathcal{M}/2$.

\begin{equation}
\Omega_{m-1,m} = \left( \frac{m-1}{m} \right) \Omega_{\rm planet}
\end{equation}

\begin{equation}
\begin{rcases}
\beta = \left( \frac{m}{m-1} \right)^{2/3} \\
\Omega = \left( \frac{m-1}{m} \right) \Omega_{\rm planet}
\end{rcases}
\text{\,\,co-rotation}
\end{equation}

\begin{align} 
\dot{e} & = +\frac{\Omega_{\rm planet}}{m} \frac{H_{\rm planet}^2}{(GM_\ast)^2} \frac{T^{\rm C}_{m-1,m}}{e M_{\rm planet}^3} \times F(p) 
\end{align}

\begin{equation}
F(p) = \frac{(1 + 0.65 p^3)^{5/6}}{(1 + 1.022p^2)^2}
\end{equation}

\begin{equation}
p = \frac{2\phi_{m-1,m}}{3\Omega^2} \left( \frac{3\Omega m}{2\beta a_{\rm planet}\nu} \right)^{2/3}
\end{equation}

\begin{align}
T^{\rm C}_{m-1,m} & = -\frac{4\pi^2 m}{3} \left( \frac{m}{m-1} \right)^2 \beta a_{\rm planet} \Omega_{\rm planet}^2 \times \nonumber \\
& \phi_{m-1,m}^2 \left( \frac{d\Sigma}{dr} + \frac{3}{2} \frac{\Sigma}{r} \right)
\end{align}

\begin{align}
\phi_{m-1,m} & = -\frac{GM_{\rm planet}e}{a_{\rm planet}} \times \nonumber \\
& \left(\frac{1}{2} - m + \frac{\beta}{2} \frac{d}{d\beta} \right) b_{1/2}^m(\beta)
\end{align}

\noindent where $b_{1/2}^m(\beta)$ is given
by \ref{eq:laplace}.

\subsection{Principal Inner Lindblad Resonances\\with $\ell = m$}
The sum runs from $m_{\rm min}=1$
to $m_{\rm max} = \mathcal{M}/2$.

\begin{equation}
\Omega_{m,m} = \Omega_{\rm planet}
\end{equation}

\begin{equation}
\begin{rcases}
\beta = \left( \frac{m}{m+1} \right)^{2/3} \\
\Omega = \left( \frac{m+1}{m} \right) \Omega_{\rm planet}
\end{rcases}
\text{\,\,inner}
\end{equation}

\begin{align} 
\dot{e} = \frac{e \Omega_{\rm planet} H_{\rm planet}^2 T^{\rm L}_{m,m}}{2(GM_\ast)^2 M_{\rm planet}^3}
\end{align}

\begin{align}
T^{\rm L}_{m,m} & = -m \pi^2 \Sigma \left(r \frac{dD}{dr} \right)^{-1} \times \nonumber \\
& \left(r \frac{d\phi_{m,m}}{dr} + \frac{2\Omega}{\Omega-\Omega_{m,m}}\phi_{m,m} \right)^2 
\end{align}

\begin{equation}
r \frac{dD}{dr} = -3\Omega^2 + 3\Omega m^2 \left(\Omega - \Omega_{\rm planet} \right) 
\end{equation}

\begin{align}
\phi_{m,m} & = -\frac{GM_{\rm planet}}{a_{\rm planet}} \left( b_{1/2}^{m} (\beta) - \beta \delta_{m,1} \right)
\end{align}

\noindent where $b_{1/2}^m(\beta)$ is given
by \ref{eq:laplace}.

\subsection{Principal Outer Lindblad Resonances\\with 
$\ell = m$}
The sum runs from $m_{\rm min}=2$
to $m_{\rm max} = \mathcal{M}/2$.

\begin{equation}
\Omega_{m,m} = \Omega_{\rm planet}
\end{equation}

\begin{equation}
\begin{rcases}
\beta = \left( \frac{m}{m-1} \right)^{2/3} \\
\Omega = \left( \frac{m-1}{m} \right) \Omega_{\rm planet}
\end{rcases}
\text{\,\,outer}
\end{equation}

\begin{align} 
\dot{e} = \frac{e \Omega_{\rm planet} H_{\rm planet}^2 T^{\rm L}_{m,m}}{2(GM_\ast)^2 M_{\rm planet}^3}
\end{align}

\begin{align}
T^{\rm L}_{m,m} & = -m \pi^2 \Sigma \left(r \frac{dD}{dr} \right)^{-1} \times \nonumber \\
& \left(r \frac{d\phi_{m,m}}{dr} + \frac{2\Omega}{\Omega-\Omega_{m,m}}\phi_{m,m} \right)^2 
\end{align}

\begin{equation}
r \frac{dD}{dr} = -3\Omega^2 + 3\Omega m^2 \left(\Omega - \Omega_{\rm planet} \right) 
\end{equation}

\begin{align}
\phi_{m,m} & = -\frac{GM_{\rm planet}}{a_{\rm planet}} b_{1/2}^{m} (\beta)
\end{align}

\noindent where $b_{1/2}^m(\beta)$ is given
by \ref{eq:laplace}.

\end{document}